\documentclass[aps,prl,epsfigure,twocolumn,longbibliography,superscriptaddress]{revtex4-1}

\usepackage{amsfonts}
\usepackage{amssymb}
\usepackage{amsmath}
\usepackage{calc}
\usepackage{subfigure}
\usepackage{graphicx}
\usepackage{epstopdf}
\usepackage{dcolumn}
\usepackage{bm}
\usepackage{color} 
\usepackage{txfonts}
\usepackage[dvipsnames]{xcolor}
\usepackage{appendix}
\usepackage[colorlinks, citecolor=blue]{hyperref} 
\usepackage{longtable,booktabs,setspace} 
\usepackage{url}
\usepackage{natbib}

\newcommand{\beq}{\begin{equation}}
\newcommand{\eeq}{\end{equation}}
\newcommand{\beqa}{\begin{eqnarray}}
\newcommand{\eeqa}{\end{eqnarray}}

\usepackage{tikz}
\usetikzlibrary{calc,decorations.markings}
\usepackage{pgfplots}

\usepackage[normalem]{ulem}   % ulem  command  to cross a text \sout{} 

\begin{document}

%\preprint{APS/123-QED}

%\textheight 24cm
%\textwidth 17cm
%\oddsidemargin 0cm
%\evensidemargin -1cm
%\topmargin -1cm
%opening
%\title{Single-shot shaped composite pulse for ultrafast high-fidelity robust quantum
%control}
%\title{Shaped pulse for ultrafast high-fidelity robust quantum control}
%

%\title{Robust control of unstable non-linear Hamiltonian systems via near-separatrix stable manifold}
\title{Robust Control of Unstable Non-linear Quantum Systems}

\author{Jing-Jun Zhu}
\affiliation{International Center of Quantum Artificial Intelligence for Science and Technology (QuArtist) \\ and Department of Physics, Shanghai University, 200444 Shanghai, China}

\author{Xi Chen}
\email{xchen@shu.edu.cn}
\affiliation{International Center of Quantum Artificial Intelligence for Science and Technology (QuArtist) \\ and Department of Physics, Shanghai University, 200444 Shanghai, China}
\affiliation{Department of Physical Chemistry, University of the Basque Country, 48080 Bilbao, Spain}

\author{Hans-Rudolf Jauslin}
\affiliation{Laboratoire Interdisciplinaire Carnot de Bourgogne, CNRS UMR 6303, Universit\'e de Bourgogne,
BP 47870, 21078 Dijon, France}

\author{St\'ephane Gu\'erin}
\email{sguerin@u-bourgogne.fr}
\affiliation{Laboratoire Interdisciplinaire Carnot de Bourgogne, CNRS UMR 6303, Universit\'e de Bourgogne,
BP 47870, 21078 Dijon, France}

\date{\today}

\begin{abstract}
Adiabatic passage is a standard tool for achieving robust transfer in quantum systems.
We show that, in the context of driven nonlinear  Hamiltonian systems, adiabatic passage becomes highly \textit{non-robust} when the target is unstable. We show this result for a generic (1:2) resonance, for which the complete transfer corresponds to a hyperbolic fixed point in the classical phase space featuring an adiabatic connectivity strongly sensitive to small perturbations of the model.
By inverse engineering, we devise high-fidelity and robust partially non-adiabatic trajectories. They localize at the approach of the target near the stable manifold of the separatrix, 
which drives the dynamics towards the target in a robust way.  
%One is based on adding of static detuning when the requested fidelity is not high. 
%This is achieved by deriving a class of exact solutions of the driven nonlinear dynamics, among which we found robust and high-fidelity solutions. 
These results can be applicable
to atom-molecule Bose-Einstein condensate conversion and to nonlinear optics. 
\end{abstract}

\maketitle
%
%

%\section{Introduction}

\textit{Introduction.-} Controlling non-linear quantum systems is central in recent applications, such as the ones involving many-particle systems in mean field \cite{GP}, e.g. for conversion of atoms into molecular Bose-Einstein condensates \cite{Mackie2000,Carr2009}, or non-linear optics \cite{Boyd,Agrawal,Longhi,Silberberg2008,NLO3}. Two- and  three-level $\Lambda$-type systems with second-order non-linearities have been shown to be non-controllable exactly in the sense that such non-linearities prevent reaching the target state exactly \cite{Tracking2013,Dorier}. However, one can approach it as closely as required,  and inverse-engineering techniques have been recently developed for that purpose \cite{Dorier}.

Besides high-fidelity requirements, an important issue is the robustness of the process, for instance with respect to imperfect knowledge of the system or to systematic deviations in experimental parameters.
 For linear problems, various robust techniques have been proposed and demonstrated, such as composite pulses \cite{Levitt:08,Genov:14}, adiabatic passage \cite{STIRAP}, optimal control \cite{OC1,OC2} or single-shot shaped pulses \cite{Daems:13,Leo2017} as a variant of shortcut to adiabaticity \cite{Chen:10,STA2},  as described in the review \cite{review12,review}.

 Their extension to non-linear dynamics is delicate since such dynamics features, in general, instabilities and non-integrability \cite{Itin2007}.
 Nonlinear quantum dynamics having the structure of a classical Hamiltonian system,
adiabatic passage techniques can be formulated for integrable systems in terms of action-angle variables of the corresponding classical Hamilton equations of motion. The adiabatic trajectory is formed by the instantaneous elliptic fixed points defined at each value of the adiabatic parameters and continuously connected to the initial condition. Obstructions to classical adiabatic passage are given by the crossing of the tracked fixed point with a separatrix, 
%in the vicinity of the instantaneous fixed point, 
which involves arbitrary small frequencies and instabilities 
\cite{Itin2007,NLAdiab2016}. Adiabatic solutions can be found in two-level systems \cite{NLAdiab2016} and in three-level systems of $\Lambda$ type \cite{Dorier}  with second- and third-order nonlinearities. Besides 
optimal control  based on Pontryagin's maximum principle \cite{Bonnard-Sugny-book,Chen16}, the use of inverse engineering techniques allows one to produce exact and controllable solutions without the need of invoking adiabatic approximations  \cite{Dorier}. 
%In $\Lambda$ three-level systems, where the (1:2) nonlinearity is for the transition involving the initial state (traditionally associated to a pump coupling), one can %derive robust solutions by imposing that the transient population in the upper state is small, in a similar way as for its linear counterpart \cite{Dorier}. Robustness %can be naturally achieved because the target state is stable in the phase space \cite{Itin2007}. 
Even non-integrability can be circumvented by appropriate design of pulse's parameters \cite{chaos}.

However, when the target state is itself unstable, e.g. associated to an hyperbolic fixed point in the classical phase space representation, as it is the case for a two-level system with a (1:2) resonance, we show the counterintuitive result that \textit{adiabatic solutions lack robustness}. The existence of robust solutions becomes then questionable. The goal of this letter is to show that one can design partially non-adiabatic trajectories, targeting an unstable state, featuring both high-fidelity and robustness.  They are built on the concept of shortcuts to adiabaticity solutions by inverse engineering adapted to non-linear dynamics.
%and in particular with respect to instability.
%The technique is applied for a non-linear model featuring a 1:2 resonance (second-oder nonlinearity) and third-order nonlinearities.
%\section{Non-linear model and the generalized Bloch sphere}

 \textit{Non-linear (1:2) resonance model and the generalized Bloch sphere.-}
We consider a nonlinear driven two-level model including a second-order nonlinearity that corresponds to a (1:2) resonance \cite{Tracking2013}:
\begin{subequations}
 \label{eq: ib1-2}
 \begin{eqnarray}
 \label{eq: ic1}
i\dot b_{1} & = &-\frac{1}{3}\left[\Delta-\Lambda_a+2\Lambda_{s}|b_{2}|^{2}\right]b_1+\frac{\Omega}{\sqrt{2}}\bar b_{1}b_{2}, \\ \label{eq: ib2}
i\dot b_{2}  & = & \frac{1}{3}\left[\Delta-\Lambda_a+2\Lambda_{s}|b_{2}|^{2}\right]b_{2}+\frac{\Omega}{2\sqrt{2}}b_{1}^2,
\end{eqnarray}
\end{subequations}
%Question du $\Delta/3$ ?
%\begin{subequations}
%\label{Model2}
%\begin{align}
%\label{Model2a}
%&i\dot c_1=\Bigl[-\frac{\Delta}{3}+\Lambda_{11}|c_1|^2
%+\Lambda_{12}|c_2|^2\Bigr]c_1 + \frac{\Omega}{\sqrt{2}}\bar c_1 c_2,\\
%\label{Model2b}
%&i\dot c_2=\frac{\Omega}{2\sqrt{2}}c_1 c_1 + \Bigl[\frac{\Delta}{3}+\Lambda_{21}|c_1|^2+\Lambda_{22}|c_2|^2\Bigr]c_2,
%\end{align}
%\end{subequations}
with the amplitude probabilities $b_1$ and $b_2$ satisfying $|b_1|^2+2|b_2|^2=1$. The time-dependent driving field couples the two states via its Rabi frequency $\Omega \equiv \Omega(t)$ (assumed positive for simplicity and without loss of generality) in a near-resonant way,  and a detuning $\Delta \equiv \Delta(t)$. The second-order nonlinearity appears in the coupling term as a (1:2) resonance and the third-order nonlinearities as diagonal terms through the coefficients $\Lambda_{a}$ and $\Lambda_{s}$ (known as Kerr terms).
%For comparison purposes, the corresponding \textit{linear} model for the variables $d_1,d_2$ is defined by
%$d_1=c_1$, $d_2=\sqrt{2}c_2$, $|d_1|^2+|d_2|^2=1$ with $\Lambda_{ij}=0$ (for all $i$ and $j$) and $\bar c_1$ ($c_1$) dropped from Eq. \eqref{Model2a} [Eq. \eqref{Model2b}] in
%the coupling term: $i\dot d_1=-\frac{\Delta}{3}d_1 + \frac{\Omega}{2} d_2$, $i\dot d_2=\frac{\Omega}{2}d_1 + \frac{\Delta}{3}d_2$.
%One can rewrite Eqs. \eqref{Model2}  by the (population preserving) transformation
%\begin{subequations}
% \begin{eqnarray}
%b_{1}&=&c_{1}e^{i\int_{t_i}^t  ds[\Lambda_{11}\vert c_1(s)\vert^2+\Lambda_{12}\vert c_2(s)\vert^2-(\Lambda_a-2\Lambda_{s}|b_{2}|^{2})/3]},\qquad\\
%b_{2}&=&c_{2}e^{2i\int_{t_i}^t  ds[\Lambda_{11}\vert c_1(s)\vert^2+\Lambda_{12}\vert c_2(s)\vert^2-2(\Lambda_a-2\Lambda_{s}|b_{2}|^{2})/3]}
%\end{eqnarray}
%\end{subequations}
%as
%in order to reveal two relevant combinations of the Kerr coefficients:
%\begin{equation}\label{eq: 222}
%\Lambda_{s}=2\Lambda_{11}+\frac{\Lambda_{22}}{2}-2\Lambda_{12}, \qquad \Lambda_{a}=2\Lambda_{11}-\Lambda_{12}.
%\end{equation}
In the language of Bose-Einstein condensation, this system \eqref{eq: ib1-2} models the transfer from atomic to molecular condensates, where $|b_1|^2$ ($|b_2|^2$) is the probability of atomic (molecular) BEC.
The term $\Lambda_a$ can be trivially compensated by a static detuning, while the $\Lambda_{s}$ term can be dynamically compensated by a time-dependent detuning, in a similar way as the one presented in \cite{Dorier} for the three-state problem.

\begin{figure}[]
	\begin{center}
		(a)\includegraphics[scale=0.9]{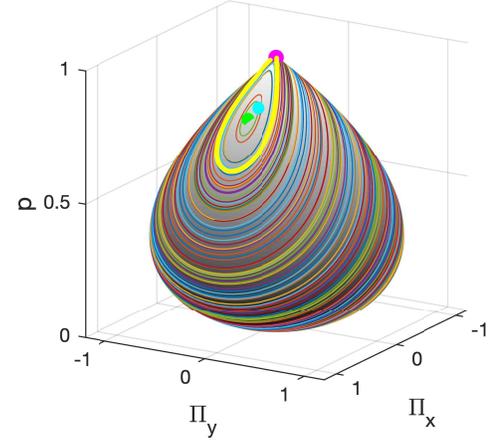}
		\\
		(b)\includegraphics[scale=0.9]{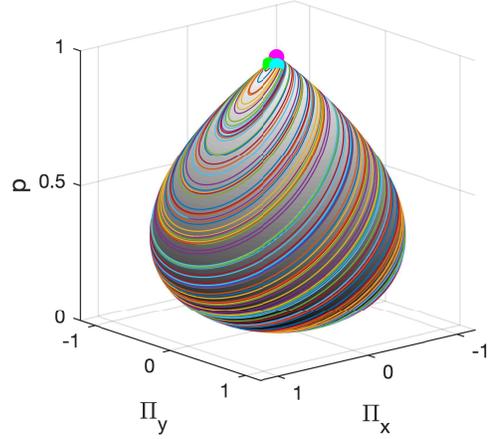}
		\caption{Portraits of the system on the generalized Bloch sphere in the late part of the dynamics ($t=1.2T$) 
		%considered in Fig. \ref{pulseshape} 
		for adiabatic tracking
		%with $T\Omega_0=5$ 
		and robust control (with parameters of Fig. \ref{Snap}), 
		%for $\beta=0$, 
		and including an additional static detuning (a) $T\Delta_0=-0.6$ and (b) $T\Delta_0=0.6$, the separatrix (thick yellow line), elliptic fixed point (green dot), actual dynamics for adiabatic tracking (cyan dot) and for robust control (magenta dot).
		%the three points are almost superposed at the scale of the figure for $T\Delta_0=1$. 
		At the chosen time, the values of the instantaneous detuning $\Delta(t)$ and of the Rabi frequency $\Omega(t)$ are almost identical in the two techniques, thus leading to the same portraits for each $\Delta_0$. \label{Portrait}}
	\end{center}
\end{figure}

Similarly to the linear counterpart, the dynamics of this non-linear system can be parametrized by three angles $\theta\in[0,\pi]$, $\alpha\in[0,2\pi[$, $\gamma\in[0,2\pi[$ as  \cite{Tracking2013,Efstathiou}:
\begin{equation}
\label{solgen}
\left[\begin{array}{cc}b_1(t)\\b_2(t)\end{array}\right]
=\left[\begin{array}{cc}\cos(\theta/2)\\
\frac{1}{\sqrt{2}}\sin(\theta/2)\,e^{-i(\alpha+\gamma)}\end{array}\right]e^{-i\gamma},
\end{equation}
%where $\gamma$ is the global phase of the wavefunction. 
%The non-linear 1:2 resonance induces the following salient differences with respect to the linear system: The normalization of the second amplitude and the relative phase, which is 
%$\alpha+\gamma$ in this representation. However the coherences involve only $\alpha$ as shown below.
The problem can be reformulated with (complex) Hamilton equations and canonical transformations into the variables $(I=\vert b_2\vert^2,\alpha)$ leads to the coordinates, respectively the opposite of the population inversion, the real and imaginary parts of the generalized coherence:
\begin{subequations}
\begin{eqnarray}
\Pi_z :=&   |c_1|^2-2|c_2|^2 &= 1-2p, \\ 
\Pi_x :=&  \null~~2 (c_1^{2}\bar c_2 + \bar c_1^{2}c_2 ) &=   2\sqrt{2}(1-p)\sqrt{p}\cos \alpha, \\
\Pi_y :=& -2i (c_1^{2}\bar c_2 - \bar c_1^{2}c_2 ) &=   2\sqrt{2}(1-p)\sqrt{p}\sin \alpha,
\end{eqnarray}
\end{subequations}
with twice state-2 population $p=2I=2\vert b_2\vert^2=\sin^2(\theta/2)$. 
%on which one can represent the dynamics. 
For convenience, one can alternatively consider the $z$-coordinate as $p$ instead of $\Pi_z$. The space phase can be reduced to a two-dimensional surface, defined as the generalized Bloch sphere, of equation $\Pi_x^2+ \Pi_y^2 = 8(1-p)^2 p, \ p\in[0,1]$,  embedded in the 3-dimensional space of coordinates $\Pi_x,\Pi_y,p$, as shown in Fig. \ref{Portrait}.

The non-linear Schr\"odinger equation leads to the following system of equations in terms of the angles and the parameters:
\begin{subequations}
\label{syst-gam-thet-phi}
\begin{eqnarray}
%\label{thetadot}
\label{popp}
%\dot\theta& = &\Omega\sin\alpha\cos(\theta/2),\hbox{ or }\dot p=\Omega(1-p)\sqrt{p}\sin\alpha=\frac{\Omega}{2\sqrt{2}} \Pi_y,\quad\ \  \\
\dot\theta& = &\Omega\sin\alpha\cos(\theta/2),\quad\hbox{ or }\quad\dot p=\Omega(1-p)\sqrt{p}\sin\alpha \\
\label{phidot}
\dot\alpha& = &\frac{\Omega}{2}\cos\alpha\frac{1-3\sin^2(\theta/2)}{\sin(\theta/2)}+\Delta-\Lambda_a+\Lambda_{s}\sin^{2}(\theta/2),\qquad\\
\label{gamdot}
\dot\gamma& = &\frac{\Omega}{2}\cos\alpha\sin(\theta/2)-\frac{1}{3}\left[\Delta-\Lambda_a+\Lambda_{s}\sin^{2}(\theta/2)\right].\qquad
\end{eqnarray}
\end{subequations}
%Equation \eqref{thetadot} can also be written in terms of $p$:
%\begin{equation}
%\label{popp}
%\dot p=\Omega (1-p)\sqrt{p}\sin\alpha.
%\end{equation}
%\crossoutg{It can be solved for any given $\Omega(t)$ and $\Delta(t)$ (keeping $\alpha(t)$ unknown):}
By Eq. \eqref{popp}, the population $p(t)$ can be expressed in terms of the angle $\alpha(t)$ as 
%\begin{equation}
%\label{solp}
$p(t)=\tanh^2\left[\int_{t_i}^t\frac{\Omega(s)}{2}\sin\alpha(s)ds \right]$,
%\end{equation}
where we have assumed an initial state $b_1(t_i)=1$ at the initial time $t_i$, i.e. $p(t_i)=0$. We consider the target of a complete population transfer $p(t_f)=1$ at the final time $t_f$. This leads to the following conclusions:

\indent  (i) The transfer probability $p$ is always lower than one. It can tend to one only in the limit of an infinite pulse area. This result is consistent with the time-optimal solution
calculated by the Pontryagin maximum principle in Ref. \cite{Chen16}.

\indent (ii) The Rabi model (for $\Delta=\Lambda_a$ and $\Lambda_s=0$ giving $\alpha=\pi/2$) gives a transfer of highest fidelity for a given pulse area $\int_{t_i}^{t_f}\Omega(s)ds$. The condition for a high-fidelity transfer is $\exp[\int_{t_i}^{t_f} \Omega(s)ds]\gg 1$, which makes the Rabi model robust with respect to the pulse area unlike its linear counterpart. We remark that this trajectory evolves on the separatrix associated to the target state $p=1$, which is a hyperbolic fixed point. 
%(see below for the detailed descritpion).

\indent  (iii) The Rabi model is however strongly sensitive to a detuning $\Delta\ne 0$ (or equivalently to a third-order nonlinearity $\Lambda_s$), since it induces oscillations in the integral of $p(t)$, 
%\eqref{solp}, 
which are more intense for a larger pulse area. This latter feature is shown below to be also the case for adiabatic dynamics.

%%%%%%%%%%%%%%%%%%%%%%%% begin  rewrite the section on non-robustness of adiabatic tracking %%%%%%%%%

\begin{figure}[]
\begin{center}
\includegraphics[scale=0.83]{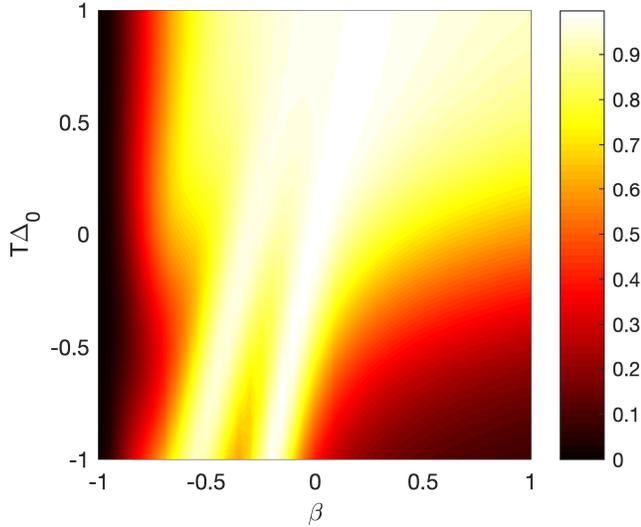}
\caption{Contour plot of the final population transfer $p(+\infty)$ for the adiabatic tracking $\Omega(t)=\Omega_{0}$ sech$(t/T)$ and $p_{\text{track}}(t)= \sin^2 [ \arctan(\sinh(t/T))/2+\pi/4 ]$ with $T\Omega_0=10$ (and $T$ the characteristic duration of the process) with respect to deviations of the detuning by a static quantity $\Delta_0$ (in units of $1/T$) and of the field amplitude by $1+\beta$.
\label{Robustadiab}}
\end{center}
\end{figure}

\textit{Dynamics in the phase space: Non-robustness of adiabatic passage.-} We  can
 limit our study for simplicity to the (1:2) resonance without third-order nonlinearities ($\Lambda_a=\Lambda_s=0$), since this system already features an unstable target.
Nonlinear adiabatic passage is expressed in terms of the dynamics of the variables $(p,\alpha)$ of the corresponding classical Hamilton equations of motion with the Hamiltonian \cite{Tracking2013}
%\begin{equation}
%\label{ham}
$h=-\Delta/3+ \Delta p/2 + (\Omega/2)(1-p)\sqrt{p}\cos\alpha$.
%\end{equation}
The adiabatic trajectory is formed by the instantaneous stable (elliptic) fixed points among the fixed points defined by $\dot p=0,\dot \alpha=0$:
\begin{equation}\label{Delta-last2}
%Adiabtic passage
\Delta=- e^{i\alpha}\frac{\Omega}{2\sqrt{p}}(1-3p),\qquad \alpha=0 \text{ or } \pi,
\end{equation}
at each value of the adiabatic parameters $\Omega\equiv\Omega(t)$ and $\Delta\equiv\Delta(t)$, and continuously connected to the initial condition $p=0$.
An adiabatic tracking trajectory is derived by imposing for instance convenient $p(t)$ and $\Omega(t)$, and using $\Delta(t)$ resulting from \eqref{Delta-last2} \cite{Tracking2013,NLAdiab2016}.
%\crossoutg{
%Figure \ref{Portrait} shows two examples of the portrait of the dynamics for various initial conditions for given (constant) values 
%$\Omega$ and $\Delta$ with $\alpha=0$, and the corresponding elliptic fixed point $(\alpha_0=0,p_0)$ with $p_0$ solution of \eqref{Delta-last2}. 
%}

The target $p=1$ is a fixed point of the dynamics, which is hyperbolic for $|\Delta/\Omega|<1$ and elliptic  for $|\Delta/\Omega|>1$. 
The number and the nature of the fixed points change as a function of $\Omega$ and $\Delta$:
%(i) For $\Omega=0$ and $\Delta=0$ all the points are fixed.\\
(i) For $\Omega=0$ and any $\Delta$ there are only two fixed points $p=0$ and $p=1$, which are both elliptic;
(ii) For $\Omega\neq0$: if $|\Delta/\Omega| < 1$ there are three fixed points: $p=1$, which is hyperbolic, and two elliptic ones. 
%located at   
%\begin{subequations}
%\label{ellipticpt}
%\begin{eqnarray}
%p_\pm &=& \left( \left[3+2\Delta^2/\Omega^2  \right]   \pm \sqrt{\left[3+2\Delta^2/\Omega^2  \right]^2 -9} \right)/9 \\
%\Pi_x  &=& \begin{cases}
%   -\frac{\Omega}{\Delta} \sqrt{2} (1-p_\pm) (1-3p_\pm),   & \text{ if~}  \Delta\neq 0, \\
%   \pm\sqrt{\frac{32}{27}}   & \text{~if~} \Delta = 0,
%\end{cases}  \\
%\Pi_y  &=& 0. 
%\end{eqnarray}
%\end{subequations}
%In particular, for $|\Delta/\Omega| =0$, $p_\pm=1/3$. 
If $|\Delta/\Omega| \geq 1$ there are two fixed points, both elliptic.

The separatrix associated to the hyperbolic fixed point is the curve of constant $h$ passing by the hyperbolic fixed point $p=1$ of equation
$( p_s-1) ( \Delta - \Omega\sqrt{p_s}\cos\alpha_s)=0$, i.e. $ \sqrt{p_s}\cos\alpha_s = \Delta/\Omega=e^{i\alpha}(1-3p_0)/(2\sqrt{p_0})$, $\alpha=0$ or $\pi$, when $ \vert\Delta/ \Omega\vert < 1$ (see Fig. 1a).
 When $|\Delta/\Omega|$ approaches $1$ from below, the separatrix collapses to a single point and $p=1$ becomes elliptic (see Fig. 1b).
 
The issue of robustness  of a typical adiabatic tracking dynamics with respect to a static detuning $\Delta_0$ and to the Rabi frequency amplitude (by multiplying it by a factor $1+\beta$) is numerically analyzed in Fig. \ref{Robustadiab}.
This shows that the fidelity dramatically decreases for negative detuning $\Delta_0$ and positive $\beta$, while it is relatively preserved on the other three quadrants. 
In what follows, we describe the dynamics in the phase space, and provide a qualitative explanation of this global lack of robustness.

\begin{figure}[]
	\begin{center}
		(a)\includegraphics[scale=0.8]{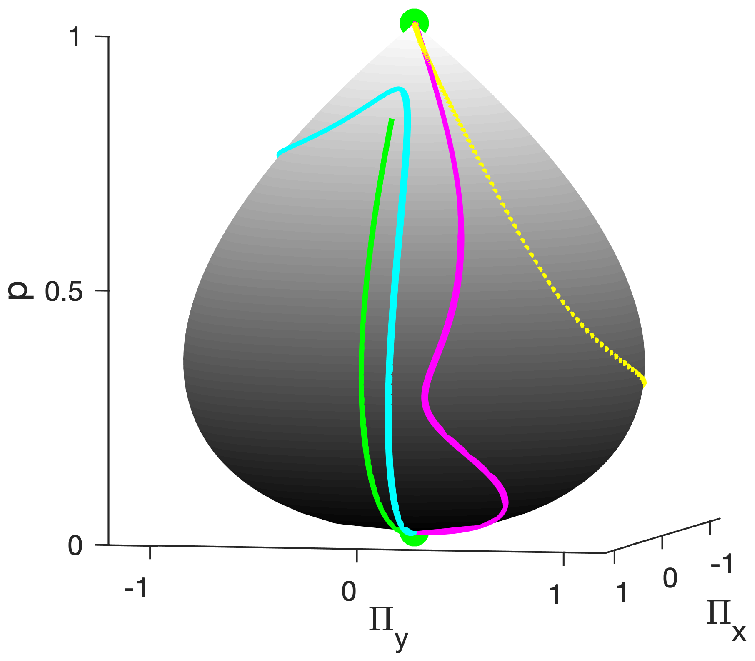}
		\\
		(b)\includegraphics[scale=0.8]{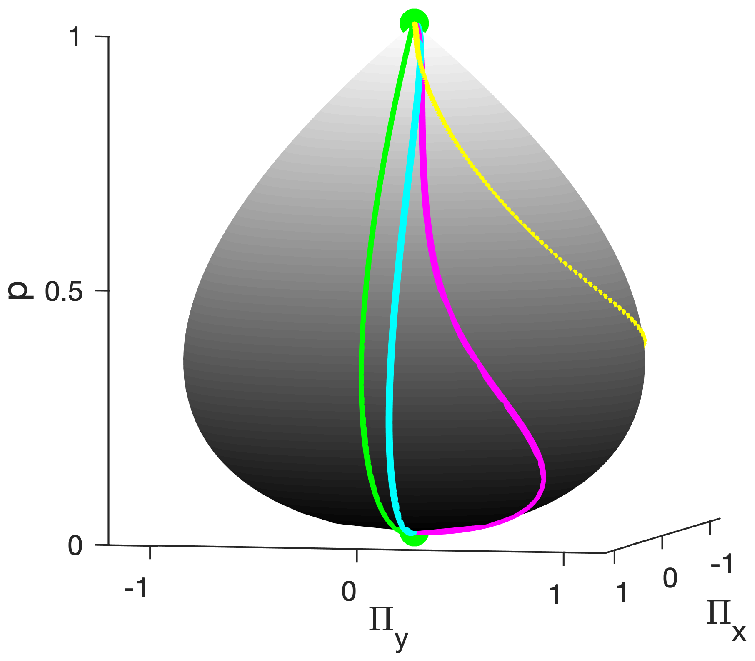}
		\caption{Trajectories with the parameters of Fig. 1 with (a) a static detuning $T\Delta_0=-0.6$ and (b) no static detuning $T\Delta_0=0$; trajectory of the instantaneous fixed points (green curves) associated to the adiabatic tracking dynamics, which connects the initial and target fixed points (green dots) in (b), but does not reach the target in (a); actual trajectory for adiabatic tracking (cyan curves) adiabatically following the green fixed point trajectory [except at the end of the dynamics in (a), when the adiabatic connectivity fails] ;  
%trajectory of the instantaneous fixed points (red curves) associated to the robust dynamics; 
actual trajectory for robust control field (magenta curves) reaching the target closely to the separatrix at the approach of the target in both cases. The separatrix (yellow) curve is made by the points of the instantaneous separatrices, each of them having the same latitude $p$ of the actual trajectory.  The four trajectories almost merge at the target in (b). 
\label{trajectoriesDm06}}
	\end{center}
\end{figure}

%The adiabatic tracking technique is based on the following observations: Since the initial state $p=0$ is a fixed point for $\Omega=0$, 
In order to reach the target $p=1$ by an adiabatic process, the trajectory must follow continuously the instantaneous elliptic fixed points that connect $p=0$ when $\Omega=0$ (intially) to $p=1$ when $\Omega/\Delta =1$ (finally) without crossing a separatrix  \cite{Itin2007,Tracking2013}, as it is shown in Fig. \ref{trajectoriesDm06}b.
The initial state $p=0$ corresponds to $\Delta/\Omega\to -\infty$ and  the target $p=1$ to $\Delta/\Omega \geq 1$. The intermediate state $p=1/3$ corresponds to  $\Delta/\Omega=0$. Thus $\Delta$ necessarily has to go through $0$. In the adiabatic tracking technique $\Delta$ is chosen such that $\Delta/\Omega \to 1$ from below at final time.
If $\Delta/\Omega = 1$ at some finite time, the elliptic fixed point collides with the hyperbolic one, and the separatrix collapses to a single point. 
%This  means that the actual trajectory crosses the separatrix right before the collapse, and thus one misses the target but of some amount, which is compatible with the fact that the 
%dynamics cannot strictly reach the target state.
If there is an additional static detuning $\Delta_0\ne0$ there are two scenarios, depending on the sign of  $\Delta_0$. We assume without loss of generality that the initial $\Delta(t_i)<0$
and thus at the approach of the target $\Delta>0$. (i) If $\Delta_0>0$, then $(\Delta+\Delta_0)/\Omega$ goes through $1$ at some finite time, then the elliptic and the hyperbolic points collide, the separatrix collapses and $p=1$ becomes elliptic (see Fig. \ref{Portrait}b). Since $\Omega\neq 0$ this implies that the actual trajectory crosses the separatrix at some earlier time (see Fig. \ref{trajectoriesDm06}b) and the adiabatic approximation is broken. However, during the crossing the flow goes into the direction of the separatrix which points toward the target, despite broken adiabatic approximation.
%If the crossing happens when $\Omega$ is quite small the actual trajectory does not move much. 
This explains the  relative robustness of the process for $\Delta_0>0$.
(ii)  If $\Delta_0<0$, 
%there is no crossing of the separatrix, but there is another effect that prevents to reach the target: 
since $[\Delta+\Delta_0]/\Omega < 1$, the elliptic fixed point stays at a finite distance from $p=1$, i.e. the elliptic fixed point never reaches the target: the adiabatic connectivity is broken (see Figs. \ref{Portrait}a and  \ref{trajectoriesDm06}a). 
%In fact the time evolution is such that the elliptic fixed point goes back toward $p=0$. Since by the  time of reversal of the mouvement of the fixed point $\Omega$ is small, the actual %trajectory does not move much and it stays at a finite distance of the target $p=1$. 
This is the main explanation of the lack of robustness with respect to a negative static detuning $\Delta_0$.
%Possible trajectories $p_{\text{track}}(t)$ with a simple pulse shape, found in \cite{Tracking2013,NLAdiab2016}, are presented as well in Fig. %\ref{pulseshape}. 
We can state a similar explanation of non-robustness of the Rabi frequency when it is multiplied by a coefficient larger than one. We can thus interpret this lack of robustness by the fact that the adiabatic connectivity is strongly sensitive to small perturbations of the model, which can be interpreted as a direct consequence of the instability of the target state.

We remark that a rough way to improve robustness is to add a static positive detuning $\Delta_s$, typically $\Delta_s=0.5/T$, in order to shift the solution towards a region of good robustness. We show below a more systematic search of a solution in this region, which is fast, robust and of high fidelity. 

\begin{figure}[t]
\begin{center}
\includegraphics[scale=0.7]{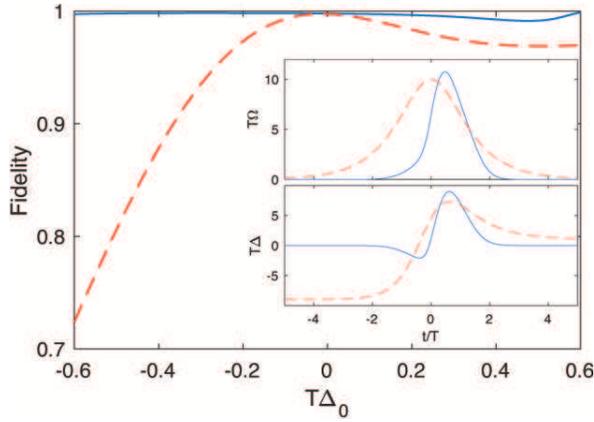}
\caption{Final transfer profile $p(+\infty)$ as a function of the static detuning $\Delta_{0}$ (in units of $1/T$) showing (i) non-robust adiabatic tracking (dashed red line) with the parameters of Fig. \ref{Robustadiab} for $\beta=0$ and (ii) robust control (solid blue line) \eqref{OmDRobust} with $C_{1}=-0.5$,  $C_{j>1}=0$, and $\epsilon=0.03$, of average transfer fidelity 0.997 in the zone of the figure.  Inset: Corresponding pulse shapes of Rabi frequency (upper frame) and detuning (lower frame). 
%for nonlinear adiabatic passage (dashed line) and robust protocol (solid line).
\label{Robust048}}
%Representation of the fixed point \eqref{fixedpoints} with $\alpha=0$ for $p=p_0=0.4,0.6,0.8,0.96$
%and the corresponding separatrix curve \eqref{separatrix}
%on the nonlinear generalized Bloch sphere.}
\label{Snap}
\end{center}
\end{figure}

\textit{Robust control.-}
We derive robust alternative solutions on the basis of reverse engineering and shortcut to adiabaticity solutions by adapting the technique developed for linear models in \cite{Daems:13}. We assume the time variation of $\theta(t)$, for instance $\theta(t)=\frac{\pi}{2}(1-\epsilon)[1+\text{erf}(t/T)]$ (where $\epsilon=0.03 >0$ is introduced in order to take into account that the solution cannot reach the target state exactly), 
%according to \eqref{solp}), 
and we define an expansion of the phase $\gamma$ as a function of $\theta$, $\tilde\gamma(\theta)\equiv\gamma(t)$, with $n$ unknown constants, $C_j$, $j=1..n$:
\begin{equation}
\label{gamma-exp}
\tilde\gamma(\theta)=\theta+C_1\sin(\theta)+C_2\sin(2\theta)+...+C_n\sin(n\theta).
\end{equation}
We determine $\alpha$ from \eqref{gamdot} and  \eqref{phidot}  by eliminating $\Delta$ and replacing $\Omega$ using \eqref{popp}, giving a differential equation for $\alpha$ as a function of $\theta$, $\tilde\alpha(\theta)\equiv\alpha(t)$:
%\begin{subequations}
%\begin{eqnarray}
%3\dot\gamma + \dot\alpha % &=& \frac{\Omega}{2} \cos\alpha \biggl[3\sin(\theta/2)
 %+ \frac{1-3\sin^2(\theta/2)}{\sin(\theta/2)} \biggr]\nonumber\\
% = \frac{\Omega}{2} \frac{\cos\alpha} {\sin(\theta/2)},
%\end{eqnarray}
%from which we replace $\Omega$ using \eqref{popp}:
%\begin{equation}
%\dot\alpha = \frac{\dot\theta}{\tan\alpha\sin\theta}  - 3\dot\gamma.
%\end{equation}
%This
\begin{equation}
\frac{d\tilde\alpha}{d\theta} = \frac{1}{\tan\tilde\alpha\sin\theta}  - 3\frac{d\tilde\gamma}{d\theta}.
\end{equation}
%This has to be solved numerically for each value of the $C_j$'s.
We remark that this equation is defined at $\theta=0$ for $\tilde\alpha(0)=\pm\pi/2$.
The field shaping is then determined from \eqref{popp} and  \eqref{gamdot}, respectively:
\begin{subequations}
\label{OmDRobust}
\begin{eqnarray}
\label{OmRobust}
\Omega(t) & = & \frac{\dot\theta}{\sin\alpha\cos(\theta/2)}, \\
\label{DRobust}
\Delta(t)& = &\frac{3}{2}\cot\alpha\tan(\theta/2)- 3\dot\gamma + \Lambda_a-\Lambda_{s}\sin^{2}(\theta/2).\qquad
\end{eqnarray}
\end{subequations}
We have to determine numerically the coefficients $C_j$'s leading to a desired robust transfer. 

%Table 1 summarizes derived coefficients with the corresponding pulse area, for which we obtain robustness as a good average fidelity in a given zone of parameters. 
%\begin{table}[h!]
%  \begin{center}
%    \label{tab:table1}
%    \begin{tabular}{c | c | c | c | c | c | c  } % <-- Alignments: 1st column left, 2nd middle and 3rd right, with vertical lines in between
%		Type & $C_{1}$ & $C_{2}$ & $C_{3}$ & Area& Fidelity & Adiabatic fidelity \\
%		\hline
%		$\Delta_{0}$ & -0.50&  &  & 8.4$\pi$ & 0.98 & 0.86 \\
%		$\beta$ & -2.46&  &  & 11.3$\pi$ & 0.975& 0.97\\
%		$\beta$ $\Delta_{0}$ & -2.16 & -0.92 &  & 11.9$\pi$ & 0.956 & 0.86 \\
%		$\beta$ $\Delta_{0}$ & -2.05& -1.00& 0.35& 12.4$\pi$ & 0.964 & 0.86 \\
%   \end{tabular}
%  \end{center}
%    \caption{Robustness with respect to detuning (referred to as type $\Delta_0$), to field amplitude (type $\beta$) or to both (type $\beta\ \Delta_0$), with the parametrizations  \eqref{gamma-exp} and the coefficients $C_{j}$, $j=1, 2, 3$, ($C_{j>3}=0$), the pulse area, and the average fidelity in the zones $-1\le T\Delta_0\le 1$, and $0.1\le\beta\le0.1$. This is compared to the average adiabatic fidelity with the parameters of Fig. 2.}
%\end{table}

Figure \ref{Snap} shows the remarkable robustness achieved with respect to the static detuning $\Delta_0$ for $C_1=0.5$ and $C_{j>1}=0$ and the corresponding pulse and detuning shapes. It surpasses the robustness of adiabatic tracking with twice lower Rabi frequency area ($5\pi$ and $10\pi$, respectively).
The robustness of this derived trajectory is analyzed in the phase space (see Fig. \ref{trajectoriesDm06}). The initial trajectory starts orthogonally to the fixed point curve since the detuning is 0 when $\Omega\ne0$. As a consequence, the adiabaticity is broken at the beginning of the process. When $\Omega$ reaches a sufficiently large value the actual dynamics becomes adiabatic, but in a region that is not close to the elliptic fixed points but rather near the stable manifold $\Pi_y>0$ of the separatrix, which drives all the trajectories in its vicinity towards the target, according to \eqref{popp}: $\dot p =\Omega \Pi_y/2\sqrt{2}$, thus in a robust way.

%Table 1 also shows that one can address robustness also with respect to Rabi frequency (or with respect to both errors at the cost of pulse-area increase) but not as efficiently as %with respect to a static detuning, since, in the case of robustness with respect to solely Rabi frequency, the robustnesses given by pulse shaping and adiabatic passage are %comparable.

One can address robustness also with respect to Rabi frequency. We obtain for $C_1=-2.12$, $C_2=-0.86$, $C_3=0.35$ (leading to the pulse area $8.6\pi$), an average efficiency of 0.972 in the zone $-0.6\le T\Delta_0\le 0.6$, $0.1\le\beta\le0.1$.

\textit{Conclusion.-}
We have shown that adiabatic passage in non-linear quantum systems is not robust when the target point is unstable due to the sensitivity to small perturbations of the adiabatic connectivity. We have developed alternative robust trajectories that circumvent the instability. The main difference is that adiabatic tracking tries to follow closely  the instantaneous fixed points, while the robust control field method operates quite far away from the fixed points near the separatrix and the stable manifold. In order to do so it breaks adiabaticity at the beginning of the process when the Rabi frequency is small.
This is versatile and applicable to stimulated Raman process for $\Lambda$-type nonlinear three-level quantum systems \cite{Dorier,chaos}, with possible applications in quantum superchemistry \cite{Superchem,Superchem2,Superchem3}. In addition,
these results can be immediately transferred to 
the other scenarios, including frequency conversion beyond the undepleted pump approximation \cite{NLO3}, nonlinear coupled waveguides \cite{waveguide},  and nonlinear Landau-Zener problem for Bose-Einstein condensate in accelerating optical lattice \cite{BECOL}.

Last but not least, the success of inverse engineering and shortcuts to adiabaticity applied for non-linear systems opens the possibility of extending shared concepts such as dynamical or adiabatic invariant, counter-diabatic driving and fast-forward scaling \cite{review12,review}.

\begin{acknowledgments}
This work was partially supported by  NSFC (11474193), SMSTC (18010500400, 18ZR1415500 and 2019SHZDZX01-ZX04), and the Program for Eastern Scholar.  
XC also acknowledges Ram\'on y Cajal program of
the Spanish MCIU (RYC-2017-22482).
SG and HRJ acknowledge additional support by the French ``Investissements d'Avenir'' programs, project ISITE-BFC / I-QUINS (contract ANR-15-IDEX-03), QUACO-PRC (Grant No. ANR-17-CE40-0007-01),
 EUR-EIPHI Graduate School (17-EURE-0002) and from the European Union's Horizon 2020 research and innovation program under the Marie Sklodowska-Curie grant agreement No. 765075 (LIMQUET).

\end{acknowledgments}

%
%This research has been conducted in the scope of the International Associated Laboratory (CNRS-France \& SCS-Armenia) IRMAS.
%We acknowledge additional support from the European Union Seventh Framework Programme through the International Cooperation ERA WIDE GA-INCO-295025-IPERA.

\end{document}